%% file: neutrino_nlo1.tex
\documentclass[a4paper,12pt] {JHEP3}  
\usepackage{epsfig,multicol,cite}

\def\as{\alpha_{\mathrm{s}}}
\def\etal{{\it et~al.}}

\def\lsim{\mathrel{\raise.3ex\hbox{$<$\kern-.75em\lower1ex\hbox{$\sim$}}}}
\def\gsim{\mathrel{\raise.3ex\hbox{$>$\kern-.75em\lower1ex\hbox{$\sim$}}}}
\def\gev{\: \rm GeV} 
 
\preprint{IMSc/2002/07/19 \\
          MRI-P-020705\\[2ex]
          {\large \tt hep-ph/0208125}}
\title{NLO corrections to ultra-high energy neutrino-nucleon
scattering, saturation and small $x$ }
\author{Rahul Basu\footnote{E-mail: rahul@imsc.ernet.in}\\
    The Institute of Mathematical Sciences,
    C.I.T. Campus, \\Taramani, Chennai 600 113, India.}
\author{Debajyoti Choudhury\footnote{E-mail: debchou@mri.ernet.in}
     {\rm and} Swapan Majhi\footnote{E-mail: swapan@mri.ernet.in}\\
    Harish-Chandra Research Institute,
    Chhatnag Road, Jhusi, Allahabad 211 019, India.}
\abstract
{We reconsider the Standard Model
interactions of ultra-high energy neutrinos with matter. The next to
leading order QCD corrections are presented for charged-current and
neutral-current processes. Contrary to popular expectations,
these corrections are found to be quite substantial, especially for very
large (anti-) neutrino energies. Hence, they need to be taken into account
in any search for new physics effects in high-energy neutrino
interactions.  In our extrapolation of the parton densities to kinematical 
regions as yet unexplored directly in terrestrial accelerators, we are guided 
by double asymptotic scaling in the large $Q^2$ and small Bjorken
$x$ region and to models of saturation in the low $Q^2$ and low $x$
regime. The sizes of the consequent uncertainties are commented upon. We
also briefly discuss some variables which are insensitive to higher
order QCD corrections and are hence suitable in any search for new
physics.}

\begin{document}

\section{Introduction}

Ultra-high energy neutrinos and their interactions continue to
attract much attention. This is in spite of the fact that no
such neutrino has been seen so far (and hence 
bounds been placed on their fluxes~\cite{flux_limits}).
Much of the continuing interest has been occasioned by the
observation, in more than one detector, of  ultra-high energy cosmic
rays (UHECR).
An interaction of such cosmic rays with either the microwave
background radiation or even the atmosphere would presumably lead
to the generation of charged pions and through their decay, to extremely
energetic neutrinos~\cite{PJ}.
Alternatively, {\em primary} ultra-high energy neutrinos themselves could
lead to UHECR, thereby avoiding the GZK bound.
A possible source for such primary neutrinos
is the decay of an
extremely massive primordial relic or even a cosmic string~\cite{BS}.
Whatever their origin, it can safely be asserted that cosmic high
energy neutrinos are inextricably linked to the very high energy cosmic
rays. The experimental detection of such neutrino fluxes is thus
expected to provide rare insight into the origin of such cosmic rays
and probably to physics beyond the SM as well.
Such observations have the additional promise of probing stellar
structures \cite{stellar_structure}, for unlike
charged particles, cosmic neutrinos do not suffer any bending due to
inter-galactic magnetic fields and hence arrive on earth
in a direct line from their source.

Consequently, various experimental efforts are being planned.
Pilot experiments,
based on the optical detection of C\v{e}renkov light emitted by the muons
created in charged current
reactions of neutrinos with nucleons either in water
or in ice, include the Antarctic Muon And Neutrino Detector Array
(AMANDA)~\cite{amanda} in the South Pole ice and the one at Lake Baikal
\cite{baikal}.
The next generation experiments using similar techniques
comprise the Neutrino Telescope and
Abyss environmental RESearch (ANTARES)~\cite{antares},
the Neutrino Experiment SouthwesT Of GReece
(NESTOR) project in the Mediterranean~\cite{nestor},
as well as ICECUBE~\cite{icecube}, the proposed kilometer scale
version of the AMANDA detector. Recently, arguments have been
forwarded in favour of
facilities based on the detection of radio pulses emanating from the
electromagnetic showers created by neutrino
interactions in ice and other materials. The primary advantage of
such a technique would be the scalability
up to an effective area of $10^4\,{\rm km}^2$
and the Radio Ice C\v{e}renkov Experiment (RICE) experiment at the
South Pole~\cite{rice} is a functioning prototype.
It has also been realized that
neutrinos can initiate horizontal Extensive Air
Showers (EAS) which could be
detected by giant ground arrays and fluorescence detectors
such as the cosmic ray Pierre Auger Project~\cite{auger-neut}.
Deeply penetrating EAS could also be detected
by observing their fluorescence light from
space based  instruments such as the Orbiting Wide-angle Light-collector
(OWL)~\cite{owl} and the
Extreme Universe Space Observatory (EUSO)~\cite{euso}. Finally, there
is the newly approved balloon experiment ANtartic Impulsive Transient
Antenna (ANITA) which will look for radio
C\v{e}renkov pulses created by ultra-high energy neutrino interactions and 
emanating from 
very large chunks of the Antarctic ice cap.  Its energy threshold is about 
$10^{18}$ eV, so it will primarily be looking for GZK neutrinos
\cite{anita}.

These experiments, taken together, are
sensitive to neutrino energies of upto $10^{11}$ GeV or so.
The actual event rates are somewhat uncertain though,
as they depend crucially on both the predicted neutrino fluxes,
as well as on the ultra-high
energy neutrino cross sections which, for want of a better method,
we may only estimate by a reasonable extrapolation
beyond the measured regime.
The interaction of UHE neutrinos with matter is through deep inelastic
scattering of the neutrinos with protons and neutrons. Over the last few
years, numerous issues with regard to the nature
of the cross section of $\nu-N$ scattering (where N is a proton or a
neutron) have gained importance and some of these are discussed in
Refs.\cite{GQRS, RSSSV, DKRS}.

Most of the discussion on $\nu-N$ scattering has been based on the
leading order (LO) expressions for neutrino nucleon scattering (i.e. $\as$
independent). The usual procedure followed has been to use the
lowest order parton level cross section and convolute it with the
LO or sometimes even next to leading order (NLO) parton
distributions. QCD corrections to the partonic cross sections have
typically been neglected, in view of the high energies involved and the
consequent small value of the strong coupling constant $\as$. While,
at first sight, such an approximation may seem appropriate,
it must be borne in mind that the consequent uncertainties may limit
the sensitivity of neutrino
telescopes~\cite{amanda,antares,nestor,icecube,rice} (and to
a smaller extent, the cosmic-ray detectors~\cite{auger-neut,owl,euso})
to physics beyond
the Standard Model. Amongst possible such scenarios, of particular
interest are theories with supersymmetry~\cite{susy}, extended
gauge or higgs sector~\cite{exotic} or, more recently, those with
a low energy gravity sector~\cite{extradim}. Perhaps, of even more
importance, are the effects on the determination of neutrino
mixing parameters~\cite{oscill}, and neutrino-tomography of the
earth's interior~\cite{tomography}. The importance of a more accurate
estimation of the neutrino interaction rates as well as their kinematical
distributions, thus, cannot be overstated.

In this paper, we explicitly calculate the ${\cal O}(\as)$ QCD corrections
to the parton model result and show that while it is not very large, it is
by
no means negligible. Moreover, we study carefully the behavior of this
correction as a function of neutrino energy and find behavior which is
not necessarily very intuitive. For example, there is a delicate interplay
between the magnitude of $\as$, the structure of the higher order
integrals and the size of the parton distributions (particularly the
gluon) in LO and NLO. This gives a non-trivial energy dependence to the
ratio of the LO and NLO cross section (which we will call the $K$-factor).
We work throughout in the
$\overline{MS}$ scheme.

Another issue which is of relevance in these energy ranges (and which,
again, has been addressed in Refs.\cite{GQRS, RSSSV, DKRS}) is the
question
of carrying out perturbative calculations at ultra low Bjorken $x$ (down
to $10^{-8}$). No data exists to help in parametrisations of parton
distribution functions at such values of $x$ and one can only be guided
in these regions by a somewhat improperly understood physical picture of
a highly dense nucleon of partons. Data from HERA stops around $x\simeq
10^{-5}$ or so and below that, some physical picture of shadowing and
saturation effects (particularly at low $Q^2$) needs to be incorporated
to understand the physics of a nucleon with a high density of partons.

We have tried to address both these issues in this paper. We have
explicitly calculated the ${\cal O}(\as)$ corrections to the partonic
cross sections and convoluted them with appropriate parton
distributions. We have also addressed the issue of extrapolation of the
partonic distributions to regions where simple DGLAP
evolution is not expected to hold.

The paper is organised as follows. In Section 2, we present a discussion
and justification for the various partonic distributions that we have
used in various parts of the $(Q^2,x)$ plane. In Section 3, we present
detailed expressions for the ${\cal O}(\as)$ corrections to the lowest
order partonic cross section for neutrino and antineutrinos scattering
against an isoscalar target. These expression are, of course available
elsewhere but for the sake of completeness and clarity, we feel it would
be
useful to present them in a form that is amenable to discussions later
in this paper. In Section 4, we present our results for LO and NLO
cross sections, both for the differential distributions $d \sigma/d \log
x$ and $d\sigma/d \log Q^2$,
as well as the total cross section. This has been done for
neutral as well as charged current cross sections. Section 5 has a short
discussion on saturation and the final section makes a few concluding
remarks.

\section{Parton Distributions}

As already mentioned briefly in the introduction, deep inelastic
scattering of UHE neutrinos is unique in the sense that it explores
extreme regions of the $(Q^2,x)$ phase space where no data from
terrestrial accelerators exist to help with the nature of the
distributions. We are therefore forced into making some assumptions
regarding the nature of these distributions, particularly in the region
of ultra low $x$ and low $Q^2$ (the so-called saturation region), as well
in the region of ultra low $x$ and very high $Q^2$, for which too, no
data exist. In what follows we shall expand a bit on this theme.

In the region between, say $50\ {\rm GeV}^2 < Q^2 < 10^5\ {\rm GeV}^2$, 
standard parametrisations are expected to work reasonably well. All
parametrisations in these regions use parton densities of the form 
$A x^{\alpha}(1-x)^{\beta} f(\sqrt{x})$ where the last function is some
polynomial in $\sqrt{x}$. 
We have checked explicitly that all three of CTEQ5
\cite{CTEQ}, MRS99 \cite{MRS} and GRV98 \cite{GRV} parametrisations 
give very similar results within this region. However, the explicit 
parametrization for either of CTEQ and MRS ditributions 
do not work below $x < 10^{-5}$ whereas
GRV98 allows $x$ values upto $10^{-9}$. For our numerical results, 
we therefore work with GRV98, except in the saturation region and the 
very large $Q^2$ and small $x$ region. 

The region above $Q^2 = 10^6\  {\rm GeV}^2$ and low $x$ ($x < 10^{-5}$) is 
beyond
the region of validity of the {\em explicit parametrisations} of GRV 
(which is valid only upto $Q^2 = 10^6\ 
{\rm GeV}^2$). One option here is to use the starting distributions for
GRV and evolve them (to LO or NLO order) into the above region. However,
in this region,  one also has the option of using an
analytical form for the distribution function - the so-called double
leading log (DLL) or Double Asymptotic Scaling (DAS) forms. In this approach,
the DGLAP equations can be solved analytically, assuming that the
evolution is driven by the splitting function $P_{gg}$. This was first
shown in \cite{dgptwz} and later developed for ultra-high energy
neutrinos in \cite{mr}. A general review with calculational details may
be found in \cite{BF}.  This region is then matched appropriately at the
boundary with GRV. 
One could also use, other than
variations of the DGLAP approach, the $\ln(1/x) $ resummation programme of
BFKL or the unified DGLAP-BFKL treatment of \cite{kms1}. The numerical 
changes are insignificant.

For the other region where the explicit GRV parametrization is not valid, 
namely large $Q^2$ ($ > 10^6 \ {\rm GeV}^2$) and moderate $x$ 
($x > 10^{-5}$), one may again evolve the parton densities appropriately 
and then use them. However, numerically, it 
is simpler to just use GRV98 with a fixed value of $Q^2 = 10^6\ {\rm GeV}^2$,
and, within this region, the error due to such an approach is negligible. 
It might seem counterintuitive that the final result depends
negligibly on which resummation scheme is used in the ultra-low $x$
and/or very high $Q^2$ scheme. 
However, as we show later in this paper, 
this simply follows from the fact 
that at these extreme ($x,Q^2$) values, the relative contribution 
in the integral (over $x$ and $Q^2$) to the total cross section is 
already very small. 

\FIGURE[!h]{\centerline{\hspace*{2em}
\epsfxsize=12cm\epsfysize=8.0cm
                     \epsfbox{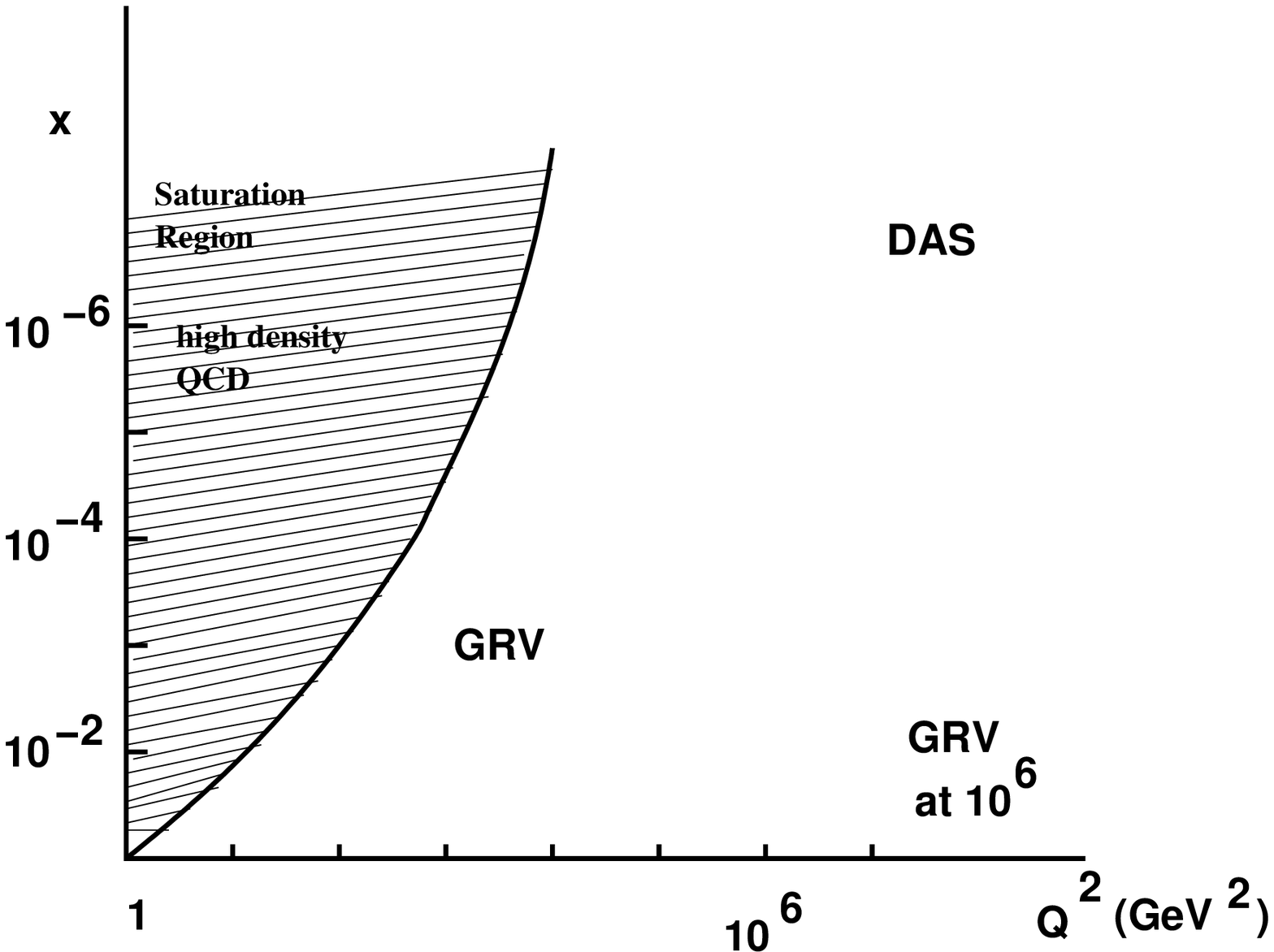}
}
\caption[]{\em Summary of the parton distributions used in various regions
	DAS stands for double asymptotic scaling (or equivalently DLL or 
	double leading log)}
\label{fig:regions}
}
The summary of the various evolution equations and their applicable
regions are shown in Fig. \ref{fig:regions}. We have used 
the leading log expressions for 
DAS; this is appropriate for matching with NLO GRV. 

Some comments regarding the saturation region are in order. In DIS at
very low $x$ and low $Q^2$, we find a high density of partons which is a
non perturbative system even though $\as$ may be small. 
There are essentially two approaches to high density QCD - the GLR
\cite{GLR} and Mueller, Qiu \cite{MQ} aproach and the effective
Lagrangian method of McLerran and Venugopalan \cite{MV}. Using these two
approaches a non-linear evolution equation has been developed by the
Tel-Aviv group and a  
review may be found in their paper \cite{GLLMNT}. We have used saturated
densities in the low $x$ and low $Q^2$ region using this approach
\cite{naftali}. Some comments on the saturation region can be found
later, in Section 5.

\section{NLO expressions for $\nu-N$ scattering}

In this section we briefly review the $O(\as)$ (NLO) expressions for
neutrino nucleon scattering. Many of these details may be found
scattered in the literature --- for example, see \cite{HvN} --- however it
is convenient to write them down here in some detail, clearly
demarcating the LO and NLO pieces. We will assume an isoscalar target
$N$ [$\equiv (p+n)/2$] as this is an excellent approximation 
to the typical neutrino detector material.

The general expression for the neutral current (NC) reaction
$$
\nu^{\!\!\!\!\!\!^{(-)}}_\mu N \rightarrow \nu^{\!\!\!\!\!\!^{(-)}}_\mu
+ anything
$$
is given by
\begin{equation}
\frac{d^2\sigma^{(\nu,{\bar \nu})N}}{dxdy}=\frac{G_F^2}{2\pi}
\Bigg(\frac{M_Z^2}{Q^2+M_Z^2} \Bigg)^2 s~\Bigg[{\rm {F_2}}(1-y) + 
{\rm{F_1}}
 xy^2 \pm {\rm{F_3}} xy(1-{1\over 2}y)\Bigg],
\label{diff_CS}
\end{equation}
where $G_F$ is the Fermi coupling, $M_Z$ the $Z$ boson mass and 
$s$ the total centre-of-mass energy squared. The quantity 
$Q^2 \ (\equiv - q^2)$ is the momentum transfer, 
the Bjorken variable $x = Q^2/(2p.q)$ with $p$ ($p^2 = M_{p}^{2}$) being 
the four momentum of the proton (or the isoscalar target) and 
$y \equiv Q^2/(x s)$.

The structure functions 
${\rm F}_i\equiv {\rm F}_i(x,Q^2)~ \big( i = 1,2,3\big) $ are given by 
the
convolution of bare parton distributions $q_i^{(0)}(y_{p})$ and the 
partonic structure function $\it{\hat F}_{i}$. In other words, 
\begin{equation}
{\rm{F_i}}(x,Q^2) = {q_{j}}^{(0)} \otimes  \it{\hat F}_{i}^{q_{j}} +
{\bar{q}_{j}}^{(0)} \otimes  \it{\hat F}_{i}^{\bar{q}_{j}} +
{g}^{(0)} \otimes  \it{\hat F}_{i}^{g},
\end{equation}
where $\otimes$ denotes the convolution defined as
\begin{equation}
{q_{j}}^{(0)} \otimes  \it{\hat F}_{i}^{q_{j}} = \int_{x}^{1} 
{dy_{p} \over y_{p}} {q_{j}}^{(0)}(y_{p}) \it{\hat F}_{i}^{q_{j}}\Big
({x \over y_{p}}\Big).
\end{equation}
Denoting the generic weak interaction vertex involving a quark $f$ by
$ \gamma^{\mu} (c^{f}_{V} \pm c^{f}_{A}  \gamma_{5})$,
the leading order (LO) partonic structure functions are given by
\begin{equation}
\it{\hat F}_{i}^{0}(z) = K_{i} \delta(1 - z)
\end{equation}
where $ 2 K_{1} = K_2 = \Big((c^{f}_{V})^{2} + 
(c^{f}_{A})^{2}\Big)$ and
$ K_{3} = 2~c^{f}_{V}~c^{f}_{A}$.
Clearly, $ K_{1,2} $ correspond to parity conserving interactions
whereas $K_{3}$ is a measure of parity violation. 
In the case of DIS, NLO  corrections implies
corrections to the parton structure functions coming from ${\cal
O}(\as)$ corrections to the partonic cross sections.

For the quark initiated  process, to order $\alpha_{s}$, 
the partonic structure functions are given by
\begin{eqnarray}
\it{\hat F}_{i}^{q}\big(z,Q^2\big)  &=& K_{i}~\Bigg \{ \delta(1 - z) + 
\frac{2 \alpha_{s}}{3 \pi} 
~\Bigg[~\zeta ~P_{qq}(z) + 
P_{qq}(z)~\ln\Big({Q^2 \over \mu^{2}} \Big) + C_{i}^{q}(z) ~\Bigg] 
\Bigg \},
\label{small_F}
\end{eqnarray}
where the first term is the zeroth-order piece and the rest 
comprise the next-to-leading order corrections. The splitting function
$P_{qq}$ is defined through
\begin{equation}
P_{qq}(z) =  {(1 + z^2) \over (1-z)_{+}} + 
{3 \over 2} \delta(1-z) \ ,
\end{equation}
where the ``plus prescription'' is defined as usual. 
The renormalisation scale $\mu$ is introduced to make the 
coupling dimensionless in $n(= 4+\epsilon)$ dimensions and 
\begin{equation}
\zeta = \left[ {2 \over n - 4} + \gamma - \ln(4 \pi) 
\right].
\end{equation}
Clearly, the collinear
singularity manifests itself in the $\zeta$ dependent term in 
eqn.(\ref{small_F}) in the limit $\epsilon\rightarrow 0$. 
And, finally, the {\em coefficient functions} $C_{i}^{q}(z)$ 
are given by
\begin{eqnarray}
C_{i}^{q}(z) &=& (1 + z^2)~
\bigg({\ln (1 - z) \over (1 - z)} \bigg)_{+} - {(1 + z^2) \over 
(1-z)}~\ln(z)  
\nonumber \\
&- & 
{3 \over 2}~{1 \over (1-z)_{+}} + 3 + 2 z - \delta(1-z) \Big( {9 
\over 2} 
+ {\pi^{2} \over 3} \Big)  + \Delta_{i}^{q},
\end{eqnarray}
with
\begin{eqnarray}
\Delta_{1}^{q} &=& - 2 z  \hspace{1.1cm}  
\Delta_{2}^{q} = 0 
\hspace{1.1cm}  \Delta_{3}^{q} = - (1 +  z), 
\end{eqnarray}

For the gluon initiated process, on the other hand, the parton 
structure functions are given by
\begin{eqnarray}
\it{\hat F}_{i}^{g}\big(z,Q^2\big)  &=& K_{i}~\frac{\alpha_{s}}{4 \pi}
	~\left[~
 \zeta ~P_{gq}(z) + P_{gq}(z)~\ln\Big({~ Q^2 \over \mu^{2}}\Big) + 
 C_{i}^{g}(z) ~\right], 
\label{small_Fg}
\end{eqnarray}
where
\begin{eqnarray}
P_{gq}(z) =  (1-z)^2 + z^2;
\end{eqnarray}
and
\begin{eqnarray}
C_{i}^{g}(z) &=& \bigg\{(1-z)^2 + z^2\bigg\}~
\ln \bigg({1 - z \over  z}\bigg) + 6~z~(1-z)  + \Delta_{i}^{g} ;
\end{eqnarray}
with
\begin{eqnarray}
\Delta_{1}^{g} = 4z(1-z) \hspace{1cm} 
\Delta_{2}^{g} =  \Delta_{3}^{g} = 0 .
\end{eqnarray}
Note that the gluonic contribution to 
$F_{3}$ vanishes identically. This feature 
follows from the $V - A$ structure of the weak
interaction whereby the quark contribution is exactly 
opposite to the anti-quark contribution. 
For the anti-quark initiated processes all the structure functions are 
the 
same except for $\it{\hat F_{3}^{q}}(z,Q^2)$ which appears with the
opposite sign in eqn.(\ref{diff_CS}). 
Absorbing the mass singularity into the bare parton distributions
in the $\overline{MS}$ scheme, the $Q^2$ dependent structure functions
can be expressed in terms of the renormalised parton distributions 
through
\begin{eqnarray}
{\rm{F_{1}}}(x,Q^2)
&=& \sum_{j=1}^{n_{f}} K^{(pc)}_{q_{j}}~ 
\Bigg\{ \Bigg({q_{j}}(x,Q^2) + {\bar{q}_{j}}(x,Q^2)\Bigg)
\nonumber\\[1ex]
&& \hspace{6em} + ~\alpha_{s}(Q^2) \Bigg[ C_{1}^{q}(z) \otimes 
({q_{j}} + {\bar{q}_{j}}) + 2 C_{1}^{g} \otimes g \Bigg] \Bigg\} \ ,
\label{F1}
\end{eqnarray}
\begin{eqnarray}
 {\rm{F_{2}}}(x,Q^2) 
&=& \sum_{j=1}^{n_{f}} K^{(pc)}_{q_{j}}~x~
\Bigg\{ \Bigg({q_{j}}(x,Q^2) + {\bar{q}_{j}}(x,Q^2)\Bigg)
\nonumber\\[1ex]
&& \hspace{6em} + ~\alpha_{s}(Q^2) \Bigg[ C_{2}^{q}(z) \otimes 
({q_{j}} + {\bar{q}_{j}}) + 2 C_{2}^{g} \otimes g \Bigg] \Bigg\} \ ,
\label{F2}
\end{eqnarray}
\begin{eqnarray}
{\rm{F_3}}(x,Q^2) &=& \sum_{j=1}^{n_{f}} K^{(pv)}_{q_{j}}~ 
\Bigg\{ \Bigg({q_{j}}(x,Q^2) - {\bar{q}_{j}}(x,Q^2)\Bigg)
\nonumber\\[1ex]
&& \hspace{6em} + ~\alpha_{s}(Q^2) \Bigg[ C_{3}^{q}(z) \otimes 
({q_{j}} - {\bar{q}_{j}})  \Bigg] \Bigg\} \ .
\label{F3}
\end{eqnarray}
In eqns.(\ref{F1} - \ref{F3}), the first term refers to the LO contribution
to ${\rm{F_i}}(x,Q^2)$ whereas the second term is the 
pure NLO contribution. The superscripts 
$pc$ and $pv$ stand for 'parity conserving' and
'parity violating' respectively.
The fact that the tree-level relation $F_2 = 2 x F_1$ is no longer valid will 
play a key role in understanding some of the results detailed in the
next section.

It is useful, at this stage, to note that for an isoscalar target we have
\begin{equation}
\begin{array}{rcl}
\sum_{j}^{n_{f}}~K^{(pc)}_{q_{j}}~\Big({\rm q}_{j} + {\bar{\rm q}}_{j} 
\Big) &=&\displaystyle
{1 \over 2} \Big( K^{(pc)}_{u} + K^{(pc)}_{d}\Big) ~\Big( u + d + 
\bar{u} + 
\bar{d} \Big)
\\[2ex]
&+&  \displaystyle
K_{d}^{(pc)} \Big(s + \bar{s} + b + \bar{b} \Big)
+  K_{u}^{(pc)} \Big(c + \bar{c} + t + \bar{t} \Big),
\\[3ex]
\sum_{j}^{n_{f}}~K^{(pv)}_{q_{j}}~\Big({{\rm q}_{j}} - {\bar{\rm 
q}}_{j} \Big) 
&=& \displaystyle
{1 \over 2} \Big( K^{(pv)}_{u} + K^{(pv)}_{d}\Big) ~\Big( u + d - 
\bar{u} - 
\bar{d} \Big),
\end{array}
\end{equation}
where  $K_{u}$ and $K_{d}$ are the couplings of the $up$ and $down$ 
type quark respectively. All parton distributions are the 
``renormalized'' parton distributions (i.e. ${\rm q}_{j} \equiv 
{\rm q}_{j}(x,Q^2)$).

The general expression for the charged current (CC) reaction 
$$
\nu^{\!\!\!\!\!\!^{(-)}}_\mu N \rightarrow \mu^{\pm} + anything,
$$
is given by
\begin{equation}
\frac{d^2\sigma^{(\nu,{\bar \nu})N}}{dxdy} =\frac{2 G_F^2}{\pi}
\Bigg(\frac{M_W^2}{Q^2+M_W^2} \Bigg)^2 s~\Bigg[{\rm F}_2(1-y) + {\rm 
F}_1 xy^2
\pm {\rm F}_3 xy(1-{1\over 2}y)\Bigg],
\end{equation}
all the notation being the same as NC. As the couplings 
$K_i$ are now flavour independent ($2 K_{1} =  K_{2} = K_{3} = {1 \over 2} $),
they have been absorbed. For the isoscalar 
target, 
the quark distributions have the form 
\begin{equation}
\begin{array}{rcl}
\sum_{j}^{n_{f}}~\Big({\rm q}_j + {\bar{\rm q}}_j \Big)  & = & \displaystyle
{1 \over 2} \Big( u + d + \bar{u} + \bar{d} \Big)
+\Big(s + b + c + t \Big),
\\[2ex]
\sum_{j}^{n_{f}}~\Big({\rm q}_j - {\bar{q}}_j \Big) & = & \displaystyle
{1 \over 2} \Big( u + d - \bar{u} - \bar{d} \Big) +
\Big(s + b - c - t \Big).
\end{array}
        \label{CC:F3}
\end{equation}

\section{Results and discussion}

Having set up the formalism, we now turn to an estimate
of the numerical size of the NLO corrections. While the exact
values of the cross sections for neutrino
and antineutrino induced processes differ, (as also for charged-current
and neutral-current interactions), one expects some similarities
in the relative size of the corrections.

\FIGURE[htb]{
\vspace*{-13.5ex}
\centerline{
\epsfxsize=17.5cm\epsfysize=12.0cm
                    \epsfbox{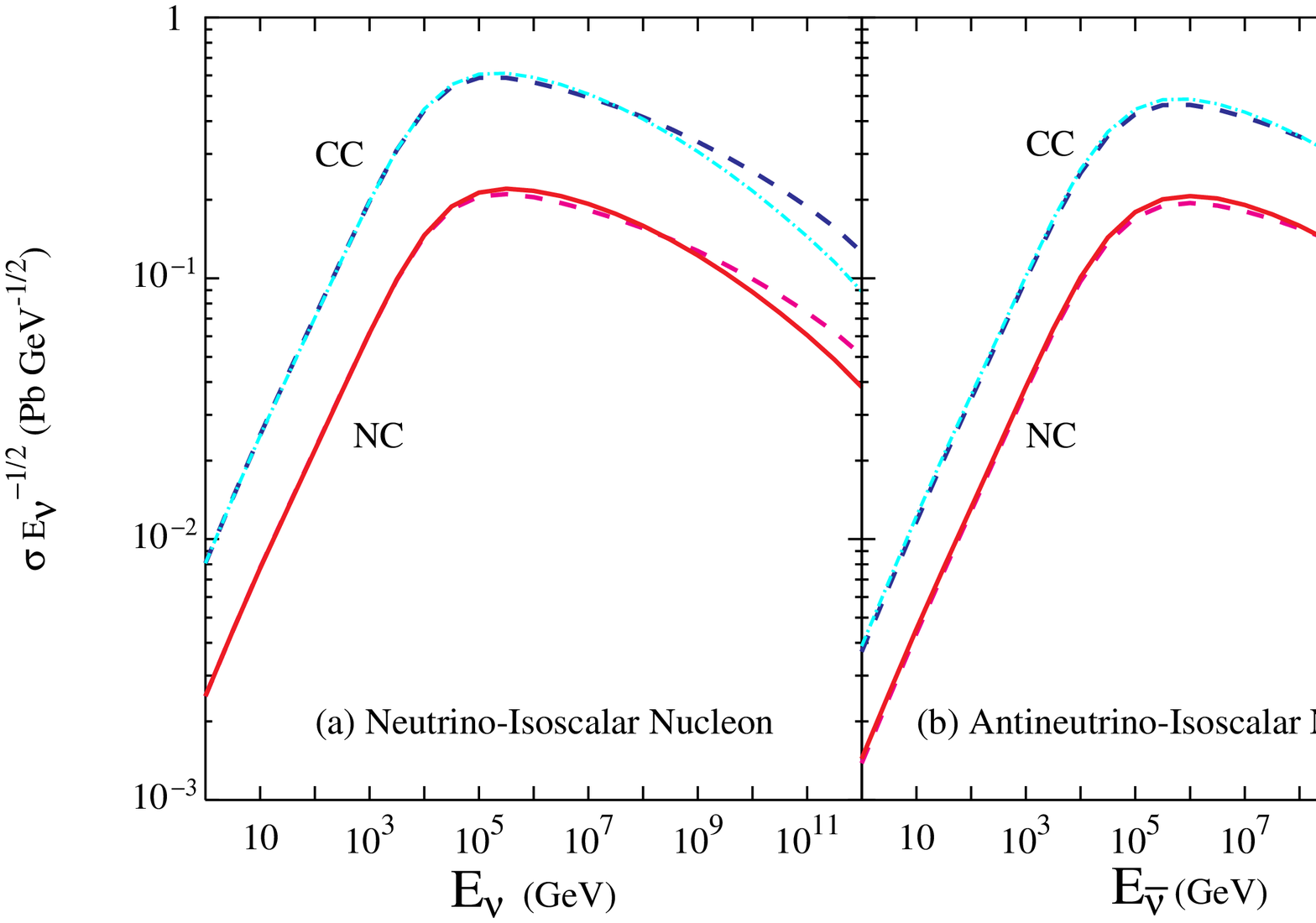}
}
\vspace*{-10ex}
\caption{\em The total NC and CC cross sections as a function of the
         (anti-)neutrino energy.
         The left and right panels correspond to neutrino and
         antineutrino scattering respectively.
         The solid (dashed) line represents the NLO (LO) cross sections.}
\label{fig:total}
}  

The total $\nu$ and $\bar\nu$ cross sections are displayed in
Fig.~\ref{fig:total}. The rationale for choosing the particular scaling
will become apparent as we progress.
While the behaviour is essentially the same as that obtained by previous
authors, the exact numbers (for the LO) differ marginally.
The change is the result of our using a
different set of parton distributions as compared to, say,
Ref.~\cite{GQRS}.
Moreover, we have taken care to use the
parton distributions appropriate for the order to which the partonic
subprocesses are being calculated in each instance.
To obtain a quantitative estimate of the differences in the total cross
sections, it is useful to express them as a function of the neutrino
energy.
While a polynomial fit for $\log \sigma$ in terms of $\log E$ is
straightforward, it is, unfortunately, not very illuminating. We
attempt, instead, a piecewise fitting of the form
\begin{equation}
  \sigma ({\rm pb}) \approx {\cal A}
                \left({E_\nu \over 1 \: {\rm GeV} } \right)^\gamma,
       \label{eq:fitparam}
\end{equation}
with the parameters ${\cal A}$ and $\gamma$ as given in
Table \ref{table:fit}.
Thus, while $\sigma$ grows almost linearly with $E_\nu$ for relatively
small neutrino energies, the growth is tempered to $\sim E_{\nu}^{0.4}$
for
larger $E_{\nu}$. The particular scaling in Fig.~\ref{fig:total}, thus,
serves to highlight the deviations at high energies from the exact
scaling relation of eqn.(\ref{eq:fitparam}). Such a behaviour for the
total cross section was found earlier, in \cite{kms}, and as demonstrated
there, it is the result of the fact that when the differential cross
section is integrated over $x$ and $Q^2$, the $Q^2$ integration is
effectively cut-off at $Q^2 \simeq M_{Z,W}^2$ coupled with the fact that
the number of valence quarks is finite.
One might wonder at this stage at
the apparent lack of unitarity reflected by the value of the
exponent $\gamma$ at high energies. In particular, in view of the rapidly
rising parton distribution functions at low $x$ predicted by all
evolution equations, the question of the total cross section saturating
the unitarity bound already at around $E_\nu \simeq 10^8$ GeV is a
serious one that is addressed, for example, in Refs.~\cite{DKRS,RSSSV}.

\begin{table}[htb]
\input{table.1}
       \caption{\em Parameters for the piecewise single power fit
               (see eqn.(\protect\ref{eq:fitparam}))
               for the cross sections in terms of the (anti-)neutrino
energy.
               }
       \label{table:fit}
\end{table}

While both Fig.~\ref{fig:total} and Table \ref{table:fit} seemingly
demonstrate that, for small neutrino energies, the difference between
the LO and the NLO cross sections is miniscule, the same certainly
cannot be said of the results at large values of $E_\nu$,
This  difference can be quantified in terms
of the $K$-factor:
\begin{equation}
       K = {\sigma_{\rm NLO} \over \sigma_{\rm LO}},
\end{equation}
which we plot in Fig.~\ref{fig:k_n}.
%
\FIGURE[!t]{
\vspace*{0ex}
\centerline{
\epsfxsize=17cm\epsfysize=12.0cm
                    \epsfbox{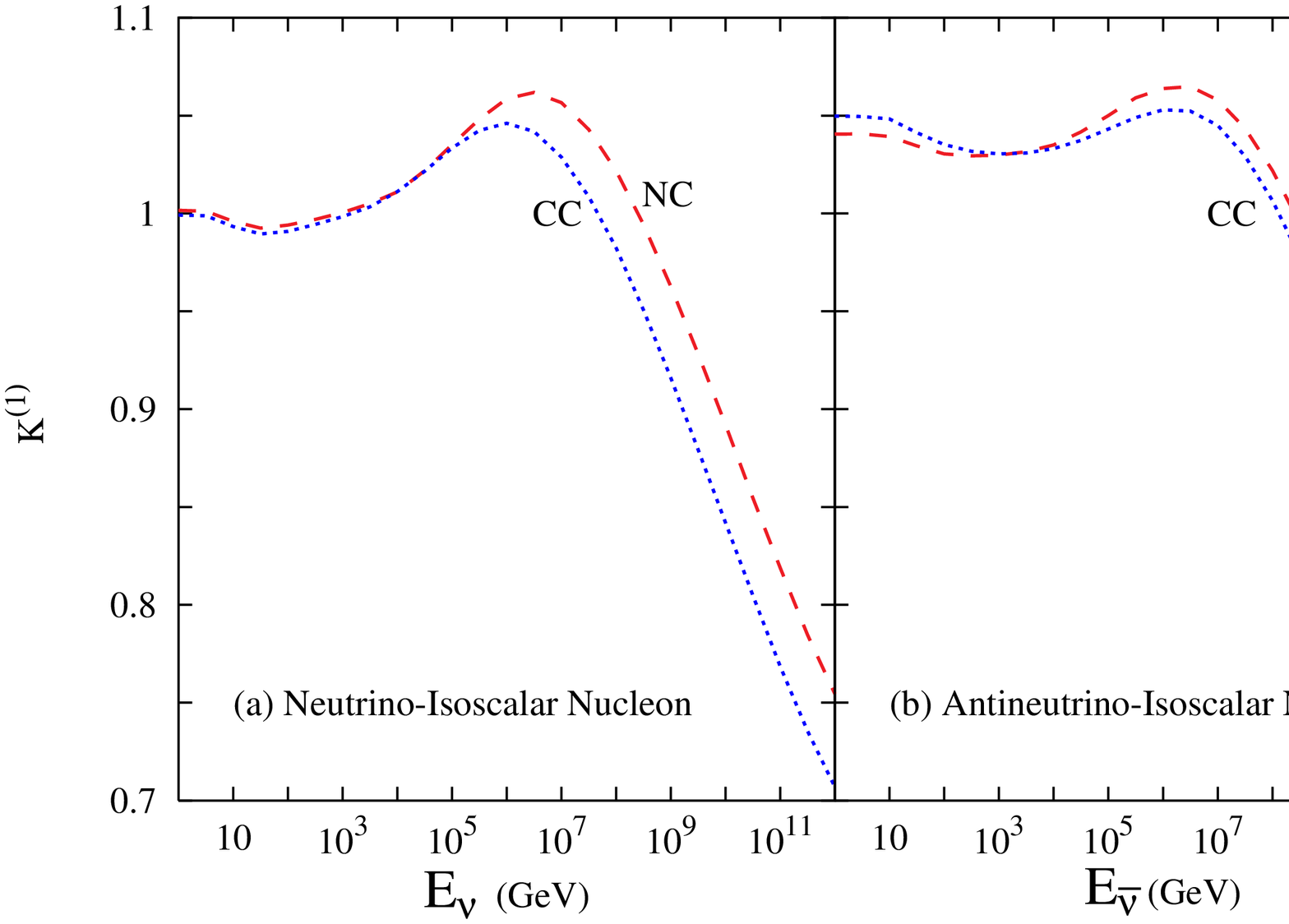}
}
\vspace*{-20ex}
\caption[]{\em The  ratio of the NLO to LO total NC and CC cross
section
         as a function of  (anti-)neutrino energy.}
\label{fig:k_n}
}

Various points are worthy of note here. As already anticipated,
for  small neutrino energies, the
$K$-factor is remarkably close to one and confirms the usual belief
that QCD corrections are unimportant. At larger energies ($E_\nu >
10^4$ GeV),
the NLO contributions start to become important and the $K$-factor
slowly
grows to its maximum ($\sim 1.08$) at about $E_\nu \sim 2 \times 10^6$
GeV.
Thereafter, the $\sigma_{\rm NLO}$ grows slower with energy than
does $\sigma_{\rm LO}$ (see Table \ref{table:fit}), resulting in a
rather strong reduction of the $K$-factor by the time one reaches
$E_\nu \simeq 10^{10}$ GeV.

\subsection{$\nu$-NC scattering}
The reasons for the afore-mentioned behaviour of the $K$-factor are
manifold and intertwined.
Hence, before we attempt an understanding of this, it is perhaps more
useful to look at the differential distributions. We begin
by concentrating on the neutral
current scattering process of neutrinos off isoscalar targets.  In
Fig.~\ref{fig:logqsq},
we show the variation with $Q^2$ for three representative neutrino
energies.
The shape of the distributions are governed by both the dynamics as
well
as the parton densities. For low neutrino energies, the maximal energy
transfer falls well short of $M_Z$ and thus the neutrino-parton
interaction
is well described in terms of a 4-fermi interaction. As is well known,
the cross section due to such an interaction Lagrangian grows with the
available center of mass energy.

On the other hand, the partons densities fall very sharply as the
fraction of the proton energy they carry approaches unity. The interplay of
these two
effects leads to the skewed bell-like shape of the distribution. As the
neutrino energy increases, so does the typical value of the
neutrino-parton
center of mass energy. Once this crosses $ s \sim (100 \gev)^2$,
the natural scale of the problem, the contribution to the
cross section is naturally dominated by virtual exchanges with
$Q^2 \sim M_Z^2$. The skewness of the distribution is therefore
removed to a large extent (see the central panel of
Fig.~\ref{fig:logqsq}). At
even larger energies, this tendency is only reinforced.
%
\FIGURE[!h]{
\centerline{\hspace*{3em}
\epsfxsize=18cm\epsfysize=8.0cm
                    \epsfbox{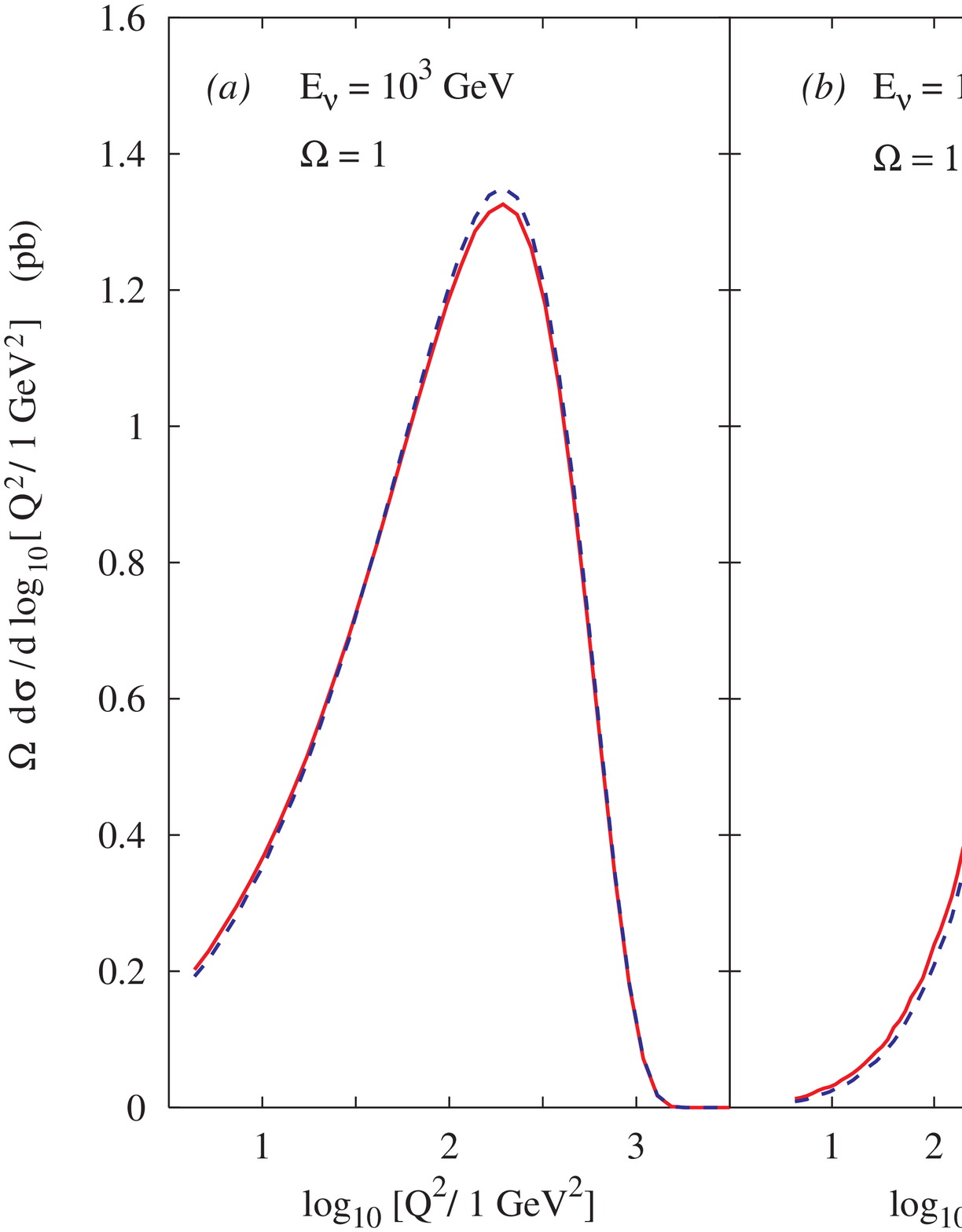}
}
\caption[]{\em The differential NC cross section wrt\  $\log(Q^2)$ for
three
         different neutrino energies. In each case, the solid (dashed)
         line give the NLO(LO) cross sections. The quantity $\Omega$
         gives the factor by which the graphs have been scaled.
         }
\label{fig:logqsq}
}

The issues discussed above can also be understood in terms of
distributions
in the Bjorken variable $x$ (see Fig.~\ref{fig:logx}). With the fast
fall of the parton densities at large $x$, it is only natural
that such regions of the phase space should contribute very little
to the cross section. This, of course, is reflected in the first two
panels of Fig.~\ref{fig:logx}. However, for very large $E_\nu$, the
situation
is changed somewhat.
Dynamics dictates that the region $Q^2 \sim M_Z^2$ (and,
hence, relatively small $x$) receives prominence in the integration. As
the
maximum in distribution is still far away from $x \lsim 1$, the
aforementioned
damping does not come into play and the fall-off is much smoother.
%
\FIGURE[!h]{
\centerline{\hspace*{3em}
\epsfxsize=18cm\epsfysize=8.0cm
                    \epsfbox{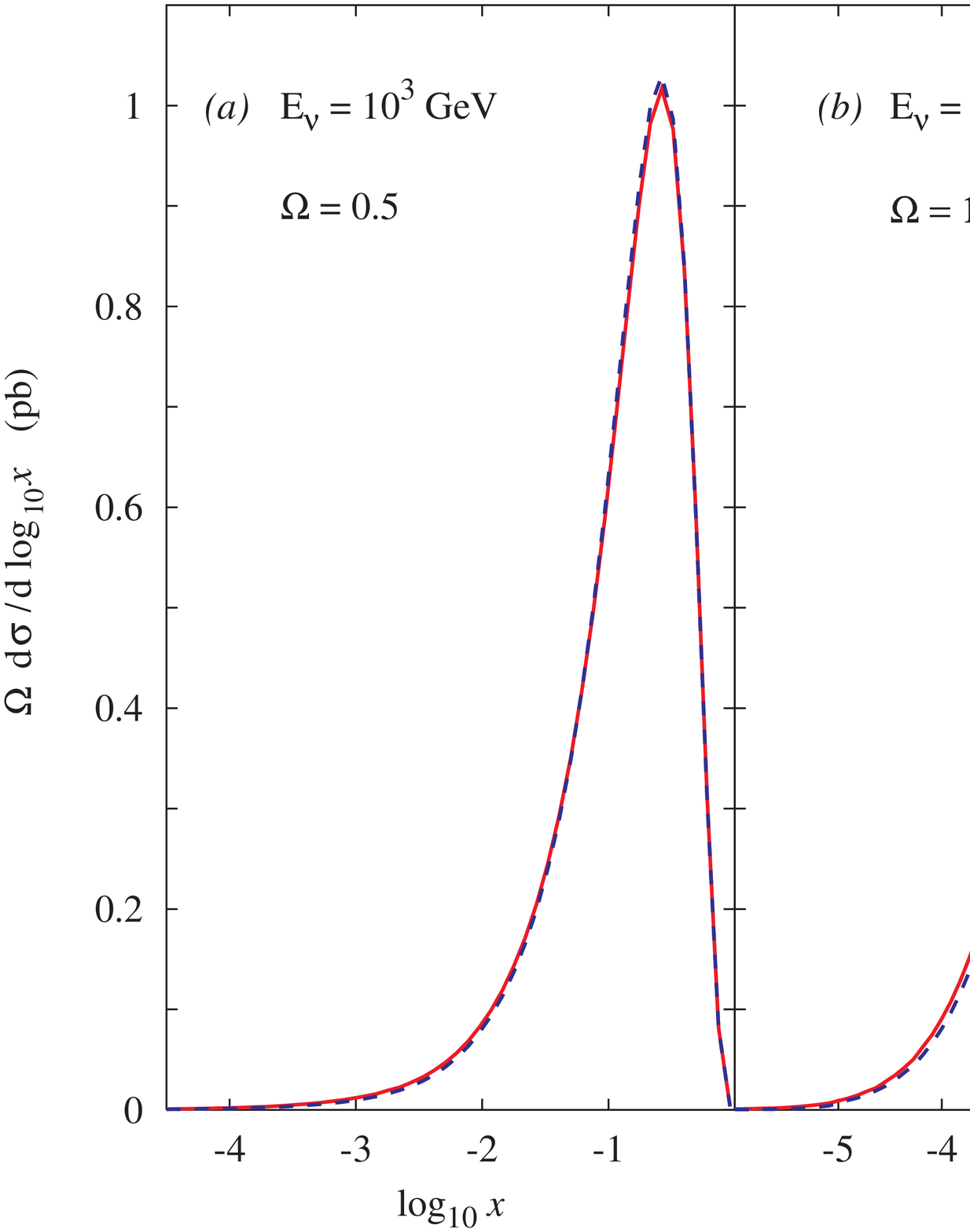}
}
\caption[]{\em As in Fig.\protect\ref{fig:logqsq}, but
           for variation in $\log(x)$ instead.
         }
\label{fig:logx}
}

With these features in mind, we now return
to the issue of the $K$-factor. To start with, let us re-examine
Fig.~\ref{fig:logqsq}, more particularly the last two panels where the
deviation is more pronounced. Even a casual glance (and this is indeed
borne
out by a quantitative test) shows that the ratio of the NLO and LO
differential cross sections is roughly constant over the significant
range of $Q^2$. The shift from unity
is as large as 25\% for extremely large $E_\nu$. As for the
$x$ distribution, a similar effect is in operation for not too large
neutrino
energies ($E_\nu \lsim 10^8 \gev$). For progressively larger
$E_\nu$'s though, the effect is concentrated more towards the peak in
${\rm d} \sigma / {\rm d} \log x$ (and may be as large as
$\sim 30\%$). We may thus conclude that, while the
NLO correction tends to be nearly independent of $Q^2$, it certainly
does
have a non-negligible variation in $x$, with the largest shifts
occurring for $x \lsim M_Z^2 / (2 M_p E_\nu)$. In other words,
were we to consider the normalized differential distributions,
the LO and NLO curves for ${\rm d} \sigma / {\rm d} \log Q^2$, for a
given
$E_\nu$,  would be virtually indistinguishable from each other, while
those for ${\rm d} \sigma / {\rm d} \log x$ would seem laterally
displaced, with the effect more pronounced for larger $E_\nu$.

The various contributions to the $K$-factor are often separated
(albeit not in a strictly gauge-invariant way) into two parts:
those arising from the differences in the LO and the NLO parton
distributions and those that are due to the changes in the
partonic cross sections. 
In Fig.~\ref{fig:kfac_tf} we display 
the effects of such a division. This would also serve to facilitate
a comparison of our results with those available in the literature.
As is seen readily, ignoring the corrections to the lowest order
partonic cross sections
but using the NLO distributions would have led to a much steeper
suppression for larger neutrino energies. The main reason for the fall
is that
while at large $E_\nu$ and hence large $s$, $x$ is small and hence the
partonic distributions are large, the rise of the distributions is
{\em steeper} for LO than for NLO. (This is, of course, a well known
fact and
arises because, at NLO, the splitting function (say $P_{qg}$) has a
$1/x$ singularity at ${\cal O}(\as)$ which needs to be stemmed by
choosing
flatter distributions at a given energy compared to the LO case.
Put differently, a stable evolution of, say, $F_2$ requires that the
NLO
steepness of $P_{qg}$ has to be compensated by a gluon density which is
less steep at NLO than at LO. It is for this reason that the starting
distributions for LO and NLO are always substantially different). In the 
literature (for example, Ref.\cite{GQRS}), it has often
been claimed that using the LO partonic cross section convoluted with
the partonic distributions in the DIS scheme is an effective method for
approximating the magnitude of the full NLO corrections. However, as
Fig.~6 clearly shows this is not the case and the 'true' NLO corrections
(wherein the NLO partonic cross section is convoluted with the NLO
distributions), gives a result very different from using just the LO
partonic cross section convoluted with either the NLO $\overline{MS}$
distribution or the distributions in the DIS scheme.

%
\FIGURE[!h]{
\vspace*{-2ex}
\centerline{
\epsfxsize=12cm\epsfysize=12.0cm
                    \epsfbox{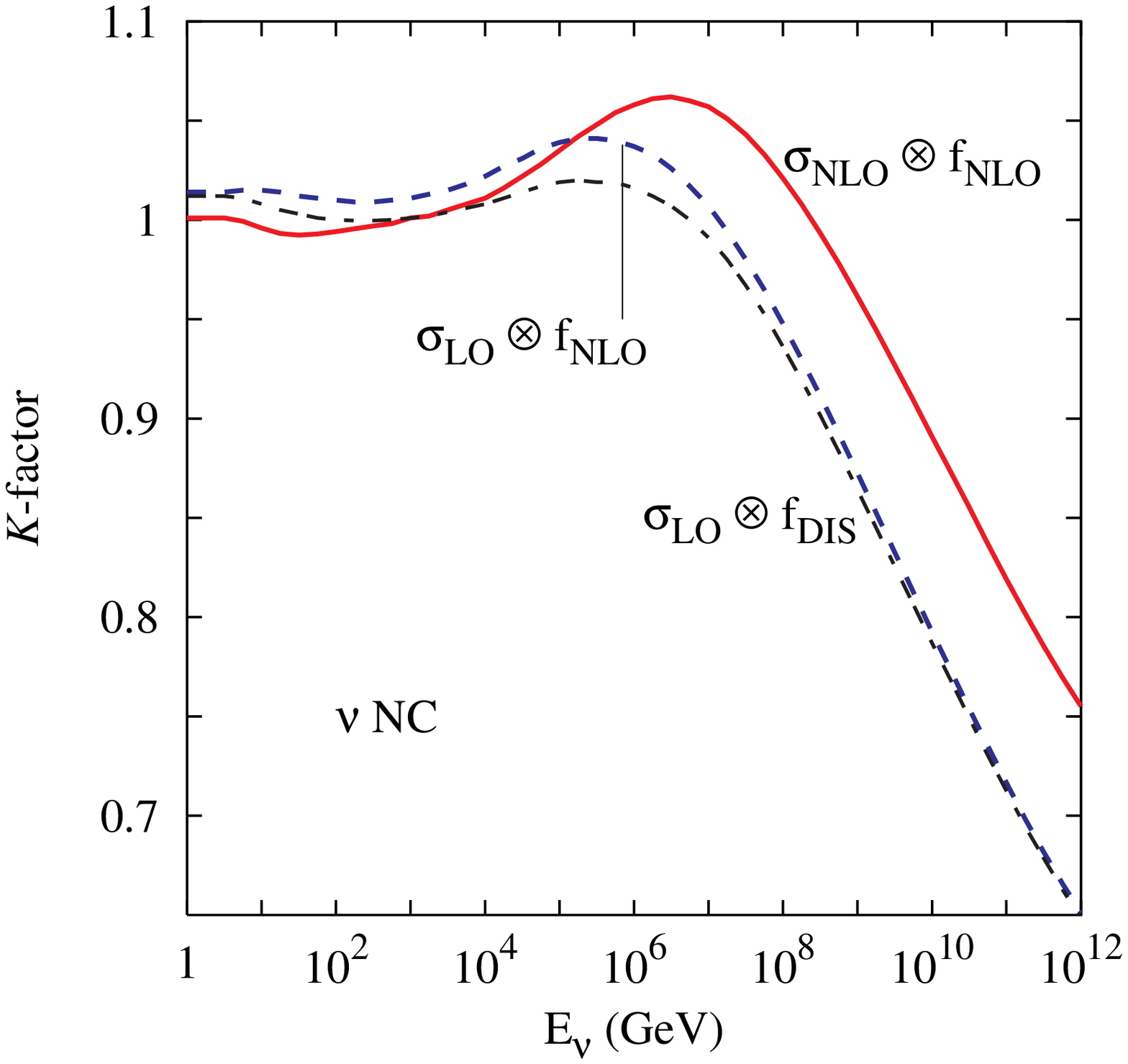}
}
\vspace*{-22ex}
\caption[]{\em A comparison of the true $K$-factor (solid line) for the
             neutrino-nucleon neutral current process and that
obtained
             if the leading order cross sections were convoluted
             with the NLO ($\overline{\rm MS}$)
             parton distributions (dashed line) or the DIS ones
             (dotted line).
         }
\label{fig:kfac_tf}
}
Having understood a decreasing $K$-factor for high neutrino eneries,
we now turn to the presence of a maximum. For sufficiently small
$E_\nu$'s,
when $x$ is not too small (see Fig.~\ref{fig:logx}), the gluon, which
is what may make the NLO large at small $x$, is still to
kick in, and the LO and NLO curves for the total cross section
start out virtually together.
As we move to larger values of $E_\nu$, and hence smaller $x$,
the gluon density itself becomes larger. Consequently,
$\sigma_{\rm NLO}$ increases faster than $\sigma_{\rm LO}$.
For even higher $E_\nu$'s, the bulk of the contribution comes from
progressively smaller $x$ and the fact of the LO distributions being
larger,
at small $x$, than the NLO ones becomes the overriding factor;
the rise in $\sigma_{\rm NLO}$ is suppressed.

Before concluding this section,  we would like to recall that
whereas the gluon is the main contributing factor in the
NLO graphs,  no gluon appears in the initial state in the LO case.
However, even for the LO, the gluon manifests itself, indirectly,
through large values of the sea quark distributions.
The inclusion of the
NLO contribution only serves to ameliorate this suppression. It
is interesting to note that the `extra' piece may contribute as much
as 10\% to the cross section and is, by no means, negligible.

\subsection{The $\bar \nu$-NC and the CC cross-sections}
Having discussed the $\nu$-NC interaction in detail, we now
comment on the other interactions that we have briefly discussed
above. As both Fig.~\ref{fig:total} and Fig.~\ref{fig:k_n} show,
the main features are quite similar and it is the small differences that
we concentrate on here. Starting with the $\bar \nu$-NC process,
note that the $K$-factor at large energies is nearly the same as that
for the corresponding $\nu$-NC interaction. While the value at
the peak ($E_{\bar \nu} \sim 10^6 \gev$) is also approximately the same,
the dip at lower energies is far more gradual. Even more striking, at
low energies  $E_{\bar \nu} \lsim 10 \gev$) is that the $K$-factor is 
discernibly
different from unity.   This difference is but an example of the
afore-mentioned subtle interplay between the parton-level cross-sections 
and the parton densities. 
At the partonic level, the only difference between the $\nu$-NC and the
$\bar \nu$-NC cross sections lies in the reversal of the sign of the
structure function $F_3$, which in turn, is controlled by the valence
quarks. At small energies, most of the contribution to the
total cross section comes from the valence quarks and hence this extra
additive piece has a significant role to play (and, indeed, is responsible
for the difference in $\nu$-NC and $\bar \nu$-NC cross sections). 
This relative shift is less pronounced for the NLO, as the $F_3$
contribution is now smaller while the combined $F_1$ and $F_2$ 
contribution is marginally larger -- again, a consequence 
of the interplay between the parton-level cross sections and the shape
of the parton densities themselves.  With the increase of the
(anti-)neutrino energy the peak in ${\rm d} \sigma / {\rm d} \log x$
shifts to progressively smaller values of $x$ (see Fig.~\ref{fig:logx}),
thereby rendering the $F_3$ contribution insignificant. 

As for the charged current processes, it is easy to see that
the NLO corrections are quite analogous to the cases already considered. 
Thus, notwithstanding the difference in the absolute value of the 
cross sections, the $K$-factor is expected to be similar, and this indeed 
is attested to by Fig.~\ref{fig:k_n}. There is a minor difference though.
Unlike the case for the NC processes, 
at high energies, the $K$-factors for the neutrino- and antineutrino-induced
CC processes are no longer the same. This has its origin in the expression
for the $F_3$ term as applied to CC (see eq.(\ref{CC:F3})). At high energies
(and, hence, low $x$), the $F_3$ contribution is now nearly proportional to 
the strange sea content in the nucleon. As the latter is, by no means, 
small at such low $x$ values, the differences in the absolute values of the 
cross sections as well as in the $K$-factors are expected.

\subsection{Observables `stable' under QCD corrections}
We have seen that all the four relevant cross sections receive significant 
corrections at the NLO level. Moreover, the corrections are not easy to 
parametrize as a function of energy. Added to this are dependencies 
on the parton distributions used. It is therefore of interest to inquire if 
there exist observables that are relatively insensitive to such corrections, 
not the least for their usefulness in searches for new physics.
One such observable is the ratio of $\sigma_\nu^{\rm CC}$ and 
$\sigma_{\bar \nu}^{\rm CC}$ (the corresponding ratio for the NC 
cross sections is not a measurable for cosmic neutrinos). In 
Fig.~\ref{fig:ratio}$a$, we exhibit this ratio as a function of 
the (anti-)neutrino energy. Since the bulk of the NLO corrections
are common to the two cross sections under discussion, the ratio, as 
expected, changes little when these corrections are included.
%
\FIGURE[!h]{
\vspace*{-20ex}
\centerline{
\epsfxsize=8cm\epsfysize=10.0cm
                    \epsfbox{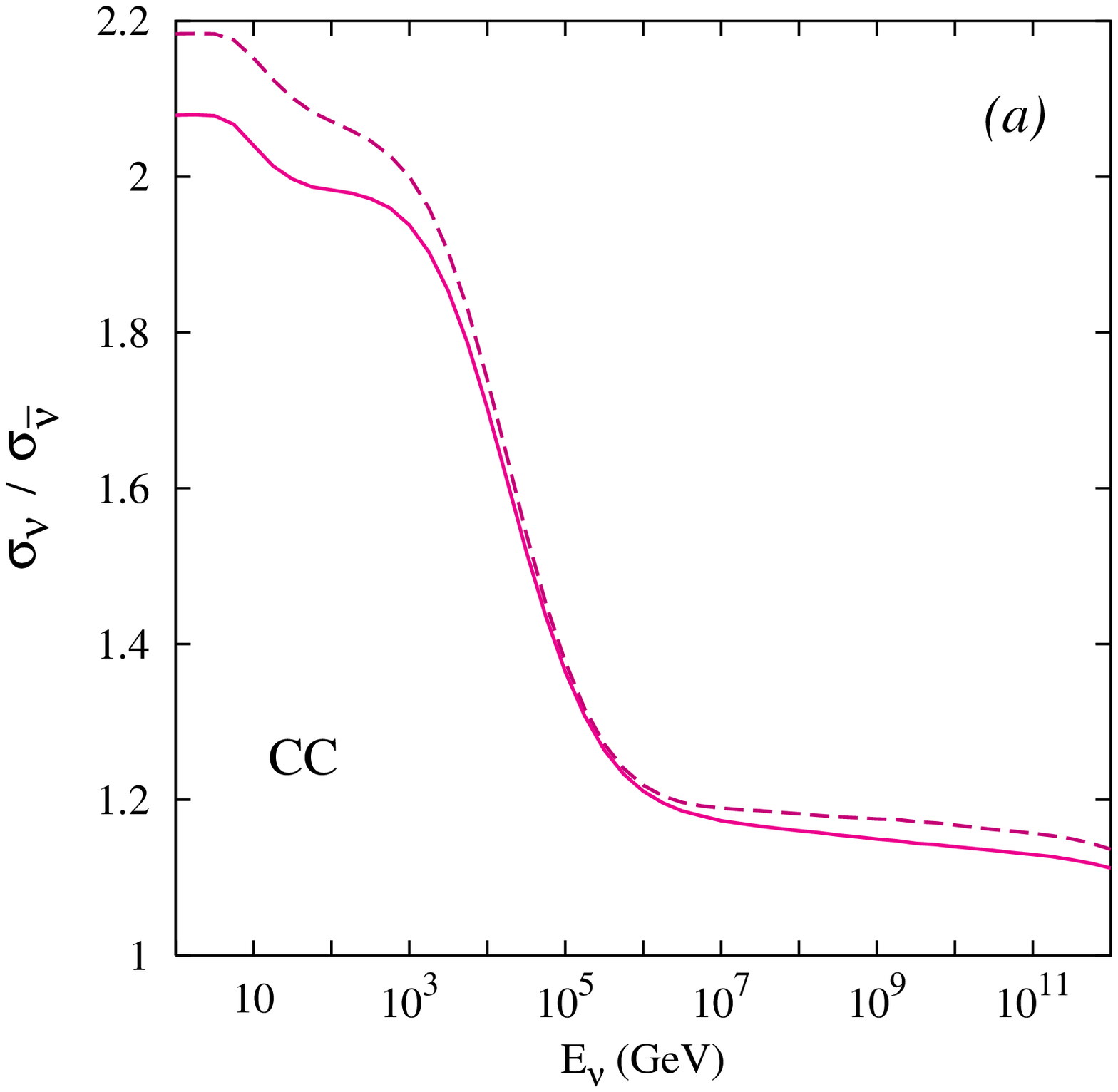}
\epsfxsize=8cm\epsfysize=10.0cm
                    \epsfbox{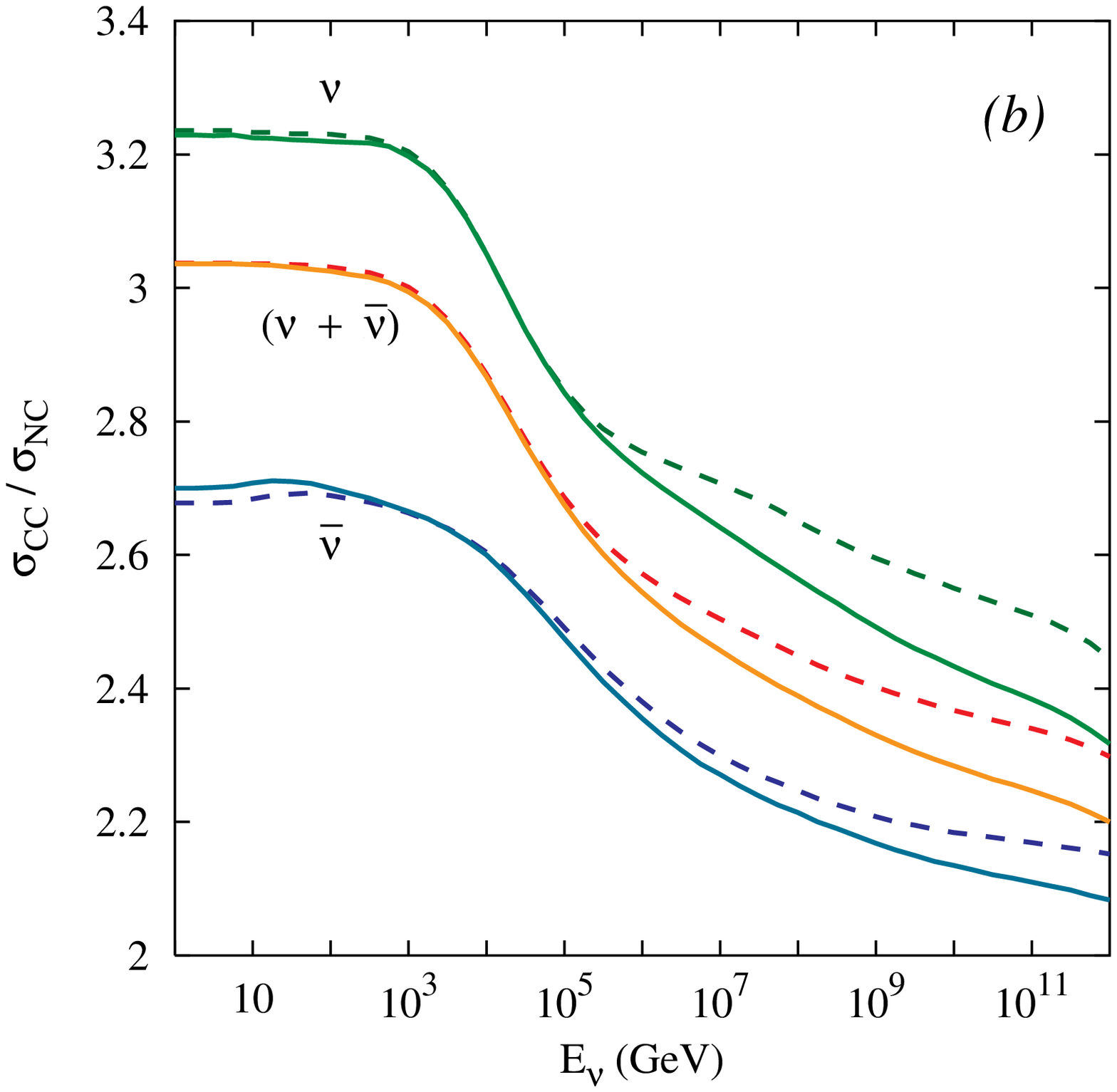}
}
\caption[]{\em {\em (a)} The ratio of the charged-current cross sections 
        for a neutrino and an antineutrino scattering off an isoscalar 
        nucleon; 
        {\em (b)} the ratio of charged-current and neutral current cross 
        sections for a neutrino, an antineutrino and a beam composed 
        of equal number of neutrinos and antineutrinos. In either case,
        the solid (dashed) lines correspond to the NLO (LO) cross sections.
         }
\label{fig:ratio}
} 

Unfortunately, in the context of an actual experiment, the ratio above 
would be measurable only in the presence of a substantial magnetic field, 
and is hence of limited use as far as a neutrino telescope is concerned. 
However, the ratio of charged current and neutral current cross sections 
is measurable, and can indeed be a good discriminator in the search for 
new physics~\cite{susy}. As Fig.~\ref{fig:ratio}$b$ shows, this ratio,
again, is far less sensitive to QCD corrections than are the 
individual cross sections.

\section{Saturation effects}
We have seen in the earlier section the complex interplay between
partonic cross sections and the evolution of partonic densities at LO and
NLO.

However, there is another issue which is of importance in this process.
At the extremely low values of $x$ that we are considering, the proton is
a dense collection of partons and saturation effects are expected to
come into play. Typically, this happens when recombination effects from $
g\ g \rightarrow g $ start having a discernible effect. Such processes are
expected
to slow down the usual rise of the gluon distribution predicted by DGLAP
evolution and is expected to take place around a scale
\begin{equation}
Q_s^2 = (1\ {\rm GeV})^2 \left(\frac{x_0}{x}\right)^\lambda \ .
\end{equation}
The parameters $\lambda$ and $x_0$ as fitted to HERA data are found
to be $\lambda=0.288$ and $x_0=3.04\times 10^{-4}$ \cite{RSSSV,GW}.
Around $x\simeq 10^{-8}$ this works out to
$Q_s^2 \simeq 20\ {\rm GeV}^2$. As is clear from the above, at larger
values of
$x$, saturation is expected to set in at even lower scales.

Keeping this in mind, we have smoothly interpolated, in the low $Q^2$ (
$\lsim 50\ {\rm GeV}^2$) and low $x$ ($\lsim 10^{-6}$) region,
from GRV to the saturation model
mentioned in the introduction \cite{GLLMNT}. All the graphs and the
analyses discussed in this paper are based on interpolation of densities
shown in Fig. 1.

While this is, of course, the consistent way of doing this, a useful
exercise would be to estimate the effect of these saturation models on
our results. To this end, we substituted the saturation model in the
low $x$ and $Q^2$ region by the standard GRV evolution and compared the
results. We find that the effect of saturation, in the total cross
section is negligible and, at its largest, is of the order of around
0.2\%.  The reason for
this is not far to seek. In the large range of $x$ and $Q^2$
values that we are spanning in our integrals, the saturation region
accounts for an exceptionally tiny part of the phase space and
consequently, has little effect on the final result. The physical
requirement of saturation is however, an important one for many
reasons, not least in stemming the rise of the distributions at ultra
low $x$ and thereby preventing the cross section from crossing the
unitarity bound. As mentioned earlier, we will not discuss this aspect here, 
but refer the reader to Refs. ~\cite{DKRS,RSSSV}.

\section{Conclusions}

In this paper, we have explicitly calculated the NLO (i.e ${\cal
O}(\as)$) corrections to ultra high energy neutrino--proton (or
isoscalar nucleon) scattering. We have identified the various regions in
the
$x$--$Q^2$ space where different parametrisations of the quark and gluon
densities are valid and discussed the contributions of each of these
regions.
In particular, we have carefully looked at the low $Q^2$ and ultra low
$x$ region and discussed the effect of saturation in this region and
found it to have a very small effect on the overall NLO behaviour. We
have also calculated the size of the NLO corrections compared with the
LO (the $K$ factor) and tried to give a qualitative explanation for the
results.

It is clear from our analysis that the higher order corrections are not
small - in fact the NLO in a certain region substantially brings {\em
down} the LO cross section and we have explained in the text why this
happens. It is clear from our analysis that NLO calculations are
necessary for UHE neutrino isoscalar scattering processes to get a
complete picture, not least for isolating the QCD effects from those
due to possible physics beyond the Standard Model.

We have also discussed briefly, observables that are insensitive to
higher order corrections. Such observables, which are 'stable' under QCD
corrections would be useful discriminators in the search for new
physics. 

At this point, one should also perhaps consider how our results would
change if one were to use other evolution equations like BFKL
\cite{bfkl} or the unified BFKL-DGLAP equation of \cite{kms1}. There
have also been attempts to use the BFKL power law rise of the structure
function and relate this power to the average inelasticity \cite{cpz}. These 
approaches are
relevant in the ultra low $x$ regions that we are discussing. However,
as has been shown in \cite{kms}, resumming leading and non-leading
$\ln(1/x)$ effects through the above equations gives neutrino cross
sections which are compatible with those obtained from the NLO DGLAP
framework, thereby reducing potential ambiguities in the extrapolation
of the cross section from the ($x,Q^2$) domain of terrestrial accelerators 
(particularly HERA) to the ultra low $x$ region that are explored by
UHE neutrinos.  

It is possible now to look at the next order in $\as$ (NNLO) to see
what effect this has at least upto values of $x \simeq 10^{-4}$ \cite{VvN}.
This is the subject of a future study.
\vskip 1cm
\noindent{\bf \large Acknowledgments}\\

One of us (RB) would like to thank Uri Maor and Eran Naftali for useful
discussions and providing us with the interpolation program for
densities in the saturation region. We would also like to thank V.
Ravindran for useful discussions.  DC thanks the Dept. of Science and
Technology, India for financial assistance under the Swarnajayanti
Fellowship grant.

\input{bib.jhep}
\end{document}

%% file: table.1
{\footnotesize
\begin{center}
\begin{tabular}{|| c | c || r | r || r | r ||}
\hline
Process & Order & 
    \multicolumn{2}{c||}{ $1 \: {\rm GeV} < E_\nu < 10^4 \: {\rm GeV}$}
                & 
    \multicolumn{2}{c ||}{ $10^7   \: {\rm GeV} < E_\nu < 10^{11} \: {\rm GeV}$}
	\\
\hline
& & \multicolumn{1}{p{10ex}}{} & \multicolumn{1}{p{10ex}||}{} 
  & \multicolumn{1}{p{10ex}}{} & \multicolumn{1}{p{10ex}||}{} \\[-5ex]
& & \multicolumn{1}{c|}{${\cal A}$} & \multicolumn{1}{c||}{$\gamma$} 
  & \multicolumn{1}{c|}{${\cal A}$} & \multicolumn{1}{c||}{$\gamma$} \\
\hline
$\nu$-NC & LO & $2.72 \times 10^{-3}$  &  0.945 &  1.17 & 0.390 \\
\cline{2-6}
     & NLO & $2.71 \times 10^{-3}$  & 0.945  &  2.06 & 0.360 \\
\hline
$\bar\nu$-NC & LO & $1.44 \times 10^{-3}$  & 0.964  &  1.16 & 0.391 \\
\cline{2-6}
     & NLO & $1.52 \times 10^{-3}$  & 0.963   &  2.06 & 0.360 \\
\hline
$\nu$-CC & LO & $8.89 \times 10^{-3}$  &  0.940 &  3.57 & 0.383 \\
\cline{2-6}
     & NLO & $8.82 \times 10^{-3}$  &  0.941 &  6.50 & 0.349 \\
\hline
$\bar\nu$-CC & LO &  $3.95 \times 10^{-3}$ & 0.961  &  2.85 & 0.386 \\
\cline{2-6}
     & NLO & $4.15 \times 10^{-3}$  & 0.960  &  5.18 & 0.353 \\
\hline
\end{tabular}
\end{center}
}

%% file: neutrino_nlo1.bbl
\begin{thebibliography}{99}

\bibitem{flux_limits} P.W.~Gorham, K.M.~Liewer and C.J.~Naudet,
 in Proc.~{\it 26th Inter. Cosmic Ray Conf.},
(Utah, 1999) [astro-ph/9906504]; \\
P.W.~Gorham et al., in Proc. {\it First Inter. 
Workshop on Radio Detection of High-Energy Particles}, 
Amer.~Inst.~of~Phys., 2001 [astro-ph/0102435] and at
{\sf http://www.physics.ucla.edu/\~{}moonemp/radhep/workshop.html}.


\bibitem{PJ}
R. J. Protheroe and P. A. Johnson, Astropart. Phys. {\bf 4} (1996) 253
[astro-ph/9506119].

\bibitem{BS}
P. Bhattacharjee and G. Sigl, Phys. Rept. {\bf 327} 
(2000) 109 [astro-ph/981101].

\bibitem{stellar_structure} 
Y.~Totsuka, Rep. Prog. Phys. {\bf 55} (1992) 377;\\
T.K.~Gaisser, F.~Halzen and T.~Stanev, Phys. Rept. {\bf 258} (1995) 173
   [hep-ph/9410384];\\
J.N.~Bahcall, \etal, Nature (London) {\bf 375} (1995) 29 [astro-ph/9503047];\\
S.~Barwick, F.~Halzen, and P.B.~Price, 
Int. J. Mod. Phys. {\bf A11} (1996) 3393 [astro-ph/9512079].

\bibitem{amanda}
G.~C.~Hill et al. (Amanda Collaboration),  astro-ph/0106064; 
{\sf http://amanda.berkeley.edu/};\\
A similar, though now defunct effort is the Deep Undersea Muon
and Neutrino Detection (DUMAND) experiment\\ 
({\sf http://dumand.phys.washington.edu/\~{}dumand/}).

\bibitem{baikal} 
Baikal Collaboration, astro-ph/9906255; \\
For general information, see {\sf http://www.ifh.de/baikal/baikalhome.html}.


\bibitem{antares} ANTARES Collaboration, astro-ph/9907432;\\
{\sf http://antares.in2p3.fr/antares/antares.html}.

\bibitem{nestor}
{\sf http://www.roma1.infn.it/nestor/nestor.html}.

\bibitem{icecube} 
AMANDA collaboration, astro-ph/9906205;\\
{\sf http://www.ps.uci.edu/\~{}icecube/workshop.html}; see

\bibitem{rice}{\sf http://kuhep4.phsx.ukans.edu/\~{}iceman/index.html}.

\bibitem{auger-neut} J.J.~Blanco-Pillado, R.A.V\'{a}zquez, and
E.~Zas, Phys.~Rev.~Lett. 78 (1997) 3614 [astro-ph/9612010]; \\
K.~S.~Capelle, J.~W.~Cronin, G.~Parente and
E.~Zas, Astropart.~Phys. 8 (1998) 321 [astro-ph/9801313]; \\
A.~Letessier-Selvon, astro-ph/0009444; \\
D.~Fargion, astro-ph/0002453, astro-ph/0101565; \\
X.~Bertou et al., astro-ph/0104452; \\
J.~L.~Feng, P.~Fisher, F.~Wilczek and T.~M.~Yu, hep-ph/0105067; \\
A.~Kusenko and T.~Weiler, hep-ph/0106071.

\bibitem{owl} J.~F.~Ormes et al., in 
	Proc.~{\it 25th International Cosmic Ray Conference},
eds.: M.~S.~Potgieter et al. (Durban, 1997) 
Vol.~5, 273; \\
Y.~Takahashi et al., in  Proc. of
International Symposium on {\it Extremely High Energy Cosmic Rays:
Astrophysics and Future Observatories}, ed. M.~Nagano (Institute
for Cosmic Ray Research, Tokyo, 1996), p.~310;\\
D.~B.~Cline and F.~W.~Stecker, OWL/AirWatch science
white paper, astro-ph/0003459;\\
see also {\sf http://lheawww.gsfc.nasa.gov/docs/gamcosray/hecr/OWL/}.

\bibitem{euso} See {\sf http://www.ifcai.pa.cnr.it/Ifcai/euso.html}.
\bibitem{anita} See {\sf http://www.phys.hawaii.edu/\~{}gorham}.

\bibitem{DKRS}
D. A. Dicus, S. Kretzer, W. W. Repko and C. Schmidt, Phys. Lett. {\bf
B514} (2001) 103 [hep-ph/0103207].

\bibitem{GQRS}
R. Gandhi, C. Quigg, M.H. Reno and I. Saracevic, Phys. Rev. {\bf D58}
(1998) 093009 [hep-ph/9807264].

\bibitem{RSSSV}
M. H. Reno, I. Saracevic, G. Sterman, M. Stratmann and W. Vogelsang,
hep-ph/0110235.

\bibitem{susy}M.~Carena, D.~Choudhury, S.~Lola and C.~Quigg, 
      Phys.Rev. {\bf D58} (1998) 095003 [hep-ph/9804380]. 

\bibitem{exotic} G.~Sigl, S.~Lee, D.N. Schramm and P.~Coppi, 
	    Phys.Lett. {\bf B392} (1997) 129 [astro-ph/9610221]; \\
 M.A.~Doncheski and R.W.~Robinett, Phys.Rev. {\bf D56} (1997) 7412
       [hep-ph/9707328];
L.~Brucher, P.~Keranen and J.~Maalampi, 
	JHEP {\bf 0105} (2001) 060 [hep-ph/0011138].


\bibitem{extradim}S.~Nussinov and R.~Shrock,  
	Phys.Rev. {\bf D59} (1999) 105002 [hep-ph/9811323; 
        Phys.Rev. {\bf D64} (2001) 047702   [hep-ph/0103043] ;\\
R.V.~Konoplich and S.G.~Rubin, JETP Lett. {\bf 72} (2000) 97 
	[astro-ph/0005225]; \\
P.~Jain, D.W.~McKay, S.~Panda and J.P.~Ralston, 
	Phys.Lett. {\bf B484} (2000) 267 [hep-ph/0001031]; \\
F.~Cornet, J.I.~Illana and M.~Masip, Phys.Rev.Lett. {\bf 86} (2001) 4235
      [hep-ph/0102065];\\
A.~Goyal, A.~Gupta and N.~Mahajan, Phys.Rev. {\bf D63} (2001) 043003 
	[hep-ph/0005030];\\
S.I.~Dutta and M.H.~Reno, hep-ph/0204218;\\
A.~Jain, P.~Jain, D.W.~McKay and J.P. Ralston, Int.J.Mod.Phys. {\bf A17} 
    (2002) 533;
A.~Ringwald and H. Tu, Phys.Lett. {\bf B525} (2002) 135 [hep-ph/0111042];\\
R.~Emparan, M.~Masip and R.~Rattazzi, Phys.Rev. {\bf D65} (2002) 064023
    [hep-ph/0109287];\\
L.~Anchordoqui and H.~Goldberg, Phys.Rev. {\bf D65} (2002) 047502
     [hep-ph/0109242];\\
J.~Alvarez-Muniz, F.~Halzen, T.~Han and D.~Hooper, Phys.Rev.Lett. 
	{\bf 88} (2002) 021301 [hep-ph/0107057];\\
M.~Kowalski, A.~Ringwald and H.~Tu, Phys.Lett. {\bf B529} (2002) 1
       [hep-ph/0201139].


\bibitem{oscill}J.~Alvarez-Muniz, F.~Halzen and D.W.~Hooper, 
	Phys.Rev. {\bf D62} (2000) 093015;\\
S.I.~Dutta, M.H.~Reno and I.~Sarcevic, 
	Phys.Rev. {\bf D62} (2000) 123001    [hep-ph/0005310];
	Phys.Rev. {\bf D64} (2001) 113015  [hep-ph/0104275]; \\ 
I.F.M.~Albuquerque and G.F.~Smoot, Phys.Rev.{\bf D64} (2001) 053008 
       [hep-ph/0102078]; \\
S.~Bottai and S.~Giurgola, astro-ph/0205325;\\
 J.F.~Beacom, P.~Crotty and E.W.~Kolb, Phys.Rev. {\bf D66} (2002) 021302
        [astro-ph/0111482].

\bibitem{tomography} P.~Jain, J.P.~Ralston and G.M.~Frichter,
Astropart.Phys. {\bf 12} (1999) 193 [hep-ph/9902206];\\ 
M. Lindner, T.~Ohlsson, R.~Tomas and W.~Winter, hep-ph/0207238. 


\bibitem{CTEQ}
 H. L. Lai, J. Huston, S. Kuhlmann, J. Morfin, F. Olness, J. F. Owens,
 J. Pumplin, W. K. Tung, Eur.Phys. J. {\bf C12} (2000) 375
 [hep-ph/9903282].

 \bibitem{MRS}
 A. D. Martin, R. G. Roberts, W. J. Stirling, R. S. Thorne, Eur.Phys.J.
 {\bf C14} (2000) 133 [hep-ph/9907231].

 \bibitem{GRV}
 M. Gl\"uck, E. Reya, and A. Vogt, Eur.Phys.J. {\bf C5} (1998) 461
[hep-ph/9806404].
\bibitem{dgptwz}
A. De Rujula, S.L. Glashow, H. David Politzer, S.B. Treiman, Frank Wilczek, 
and A. Zee, Phys.Rev. {\bf D10} (1974) 1649  
\bibitem{mr}
Douglas W. McKay and J. P. Ralston, Phys. Lett. {\bf B167} (1986) 103;
G. M. Frichter, Douglas W. McKay and J. P. Ralston, Phys. Rev. Lett.
{\bf 74} (1995) 1508, erratum {\em ibid.} {\bf 77} (1996)
4107.[hep-ph/9409433]

\bibitem{BF}
 For a review, see, R. Ball and S. Forte, hep-ph/9512208, published in
Acta Phys.Polon. B26 (1995) 2097.

 \bibitem{GLR}
 L. V. Gribov, E. M. Levin, M. G. Ryskin, Phys. Rep. {\bf 100} (1983) 1.

 \bibitem{MQ}
 A. H. Mueller and J. Qiu, Nucl. Phys. {\bf B 268} (1986) 427.

 \bibitem{MV}
 L. McLerran and R. Venugopalan, Phys. Rev. {\bf D49} (1994) 2233, 3352;
 {\bf D50} (1994) 2225; {\bf D59} (1999) 094002;
 [hep-ph/9309289,9311205,9402335,9809427] 

 \bibitem{GLLMNT}
 E. Gotsman, E. Levin, M. Lublinsky, U. Maor, E. Naftali, K. Tuchin,
 hep-ph/0101344.
 \bibitem{naftali}
 Eran Naftali, private communication.
 \bibitem{kms}
 J. Kwiecinski, A. D. Martin and A. M. Stasto, Phys. Rev. {\bf D59}
 (1999) 093002.
 \bibitem{GW}
 K. Golec-Biernat and M. Wusthoff, Phys. Rev. {\bf D59} (1999) 014017;
 {\bf D60} (1999) 114023.
\bibitem{bfkl}
E. A. Kuraev, L. N. Lipatov and V. S. Fadin, Zh. Eksp. Teor. Fiz. {\bf
72}, (1977) 373; Ya. Ya. Balitzkij and L. N. Lipatov, Yad. Fiz. {\bf 28}
(1978) 1597; J. B. Bronzan and R. L. Sugar, Phys. Rev. {\bf D17} (1978)
585; T. Jaroszewicz, Acta Phys. Pol. {\bf B11} (1980) 965.
\bibitem{kms1}
J. Kwiecinski, A. D. Martin and A. M. Statsto, Phys. Rev. {\bf D56}
(1997) 3991; Acta Phys. Pol. {\bf B28} (1997) 2577.
\bibitem{cpz}
J. A. Castro-Pena, G. Parente and E. Zas, Phys. Lett. {\bf B500} (2001)
125.
 \bibitem{HvN}
B. Humpert and W. L. van Neerven,  Nucl. Phys. {\bf B184}, (1981) 225.

\bibitem{VvN}
See, for example, W. L. van Neerven and A. Vogt, J.Phys. G28, (2002) 727
[hep-ph/0107194] and references cited therein. 

\end{thebibliography}
